# Time-magnified photon counting with a 550-fs resolution

BOWEN LI,[1] JAN BARTOS, YIJUN XIE, AND SHU-WEI HUANG[*]

*Department of Electrical, Computer, and Energy Engineering, University of Colorado, Boulder, Colorado, US*
[1] *Email: Bowen.Li@colorado.edu*
*Corresponding author: Shu-Wei.Huang@colorado.edu*

**Time-resolved photon-counting plays an indispensable role in precision metrology in both classical and quantum regimes. In particular, time-correlated single-photon counting (TCSPC) has been the key enabling technology for applications such as low-light fluorescence lifetime spectroscopy and photon counting time-of-flight (ToF) 3D imaging. However, state-of-the-art TCSPC single-photon timing resolution (SPTR) is limited in the range of 10-100 ps by the available single-photon detector technology. In this paper, we experimentally demonstrate a time-magnified TCSPC (TM-TCSPC) that achieves an unprecedentedly short SPTR of 550 fs for the first time with an off-the-shelf single-photon detector. The TM-TCSPC can resolve ultrashort pulses with a 130-fs pulsewidth difference at a 22-fs accuracy. When applied to photon counting ToF 3D imaging, the TM-TCSPC greatly suppresses the range walk error that limits all photon counting ToF 3D imaging systems by 99.2 % (130 times) and thus provides unprecedentedly high depth measurement accuracy and precision of 26 μm and 3 μm, respectively.**

Time-resolved photon counting plays an indispensable role in precision metrology in both classical and quantum regimes. Therein, time-correlated single-photon counting (TCSPC) [1] has been the key enabling technology for applications such as low-light fluorescence lifetime spectroscopy and microscopy [2, 3], time-gated Raman spectroscopy [4], photon counting time-of-flight (ToF) 3D imaging [5-7], light-in-flight imaging [8], and computational diffuse optical tomography [9]. For all these applications, the most important figure-of-merit is the single-photon timing resolution (SPTR) that directly affects the measurement accuracy and precision on important parameters such as the fluorescence-decay lifetime [2, 3], Raman spectral resolution [4], ToF distance [5-7], and spatial resolution [8, 9]. The TCSPC SPTR is currently limited by the available single-photon detector technology. Photomultiplier tubes (PMT), despite their broad spectral coverage, typically can only provide an SPTR larger than 100 ps [10]. Besides, the application of these vacuum-based devices in single-photon counting is limited by their fragility, intrinsic deterioration with usage, and bulkiness. Meanwhile, superconducting nanowire single-photon detectors (SNSPDs) have achieved superior SPTR in the sub-10-ps range [11, 12]. However, the requirement of cryogenic cooling significantly increases the system complexity and cost. In comparison, single-photon avalanche diodes (SPADs) are semiconductor devices operating at room or moderate temperature, and their CMOS compatibility enables integration into 2D detector arrays [13], which makes them a popular choice for various applications mentioned above. Nevertheless, their SPTR is still limited to tens-of-ps level for the near-infrared wavelength regime even with a customized fabrication process [14, 15]. To bridge the gap between the application needs and the TCSPC performances, an all-optical signal processing technique that can enhance the state-of-the-art SPTR by orders of magnitude is especially beneficial for many challenging applications including the study of ultrafast charge transfer and fluorescent decay dynamics [16, 17].

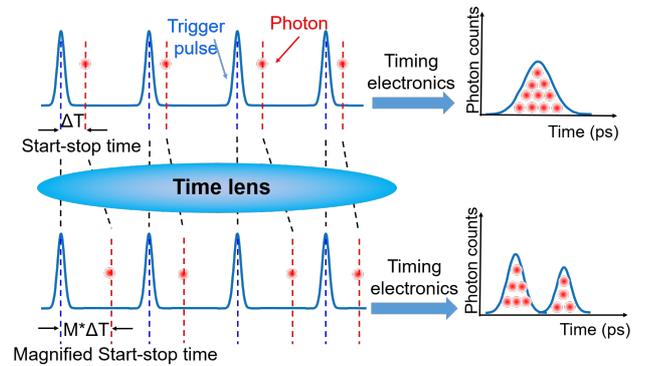

Fig. 1. Schematic diagram of TM-TCSPC. By temporally magnifying the start-stop time before detection, higher SPTR can be achieved with the same single photon detector and timing electronics.

In this paper, we demonstrate a time-magnified TCSPC (TM-TCSPC) that achieves an unprecedentedly short SPTR of 550 fs for the first time using an off-the-shelf single photon detector. The key component is a quantum temporal magnifier using a low-noise high-efficiency fiber parametric time lens [18, 19] based on four-wave mixing Bragg scattering (FWM-BS) [20, 21]. A temporal magnification of 130 with a 97% photon conversion efficiency has been achieved while maintaining the quantum coherence of the signal under test (SUT). Detection sensitivity of -95 dBm (0.03 photons per pulse), limited by the spontaneous Raman scattering noise, is possible and allows efficient processing and characterization of quantum-level SUT. In the first application, the TM-TCSPC can resolve ultrashort pulses with a 130-fs pulsewidth difference at a 22-fs accuracy. When applied to photon counting ToF 3D imaging, the TM-TCSPC greatly suppresses the range walk error (RWE) that limits all photon counting ToF 3D imaging systems by 99.2 % (130 times) and thus provides unprecedentedly high depth measurement accuracy and precision of 26 μm and 3 μm, respectively. The TM-TCSPC is a promising solution for photon counting at the femtosecond regime that will also benefit other research fields such as low-light fluorescence lifetime spectroscopy and microscopy, time-gated Raman spectroscopy, light-in-flight imaging, and computational diffuse optical tomography.

The schematic diagram is shown in Fig. 1. In conventional TCSPC (upper row of Fig. 1), the start-stop time between the excitation pulse and the emission photon is registered and logged by the TCSPC timing electronics. A TCSPC timing histogram of photon arrival time referenced to the excitation pulse is built up by repeated measurements. Provided the probability of registering more than one photon per cycle is low, the TCSPC timing histogram depicts the time-resolved intensity profile of the SUT at the quantum level. Depending on the choice of single-photon detectors, the TCSPC time resolution is in the range of 10-100 ps. In the TM-TCSPC system (lower row of Fig. 1), the quantum-level SUT is first temporally magnified before being characterized by the subsequent TCSPC system. The SPTR of the TM-TCSPC is thus significantly improved by the temporal magnification ratio, reaching the femtosecond regime for the first time. To implement the temporal magnifier that preserves the SUT quantum coherence, a fiber parametric time lens based on FWM-BS is developed. FWM-BS is advantageous for quantum applications due to its noiseless nature and near-unity conversion efficiency [20-22]. Moreover, the flexibility of choosing pump wavelength also enables processing quantum-level SUT over a large wavelength range [23].

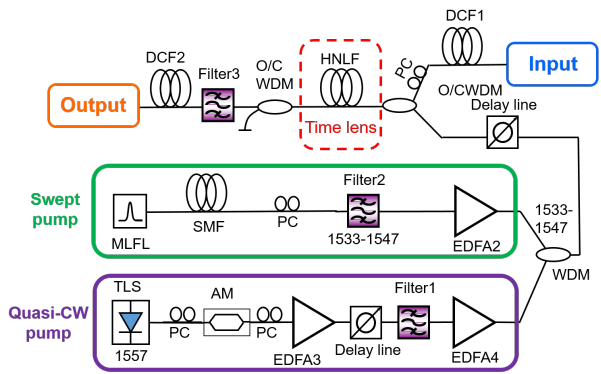

Fig. 2. Experimental set-up of the temporal magnifier using a fiber parametric time lens based on FWM-BS that is advantageous for quantum applications due to its noiseless nature and near-unity conversion efficiency. PC, polarization controller.

The experimental set-up is shown in Fig. 2. The fiber parametric time lens was implemented using a spool of 30-m highly nonlinear fiber (HNLF) and two optical pumps. First, a swept pump was generated by chirping a 100 MHz mode-locked erbium-doped fiber laser (MLFL) through 680-m single-mode fiber (SMF) to obtain a pump dispersion of -15.12 ps$^2$. It was then band-pass filtered at 1540 nm at a bandwidth of 14 nm and subsequently amplified by an erbium-doped fiber amplifier (EDFA). The swept pump pulse duration is stretched to 165 ps, defining the aperture of the time lens. At the same time, a quasi-continuous-wave (CW) pump was generated from a tunable laser source (TLS) operating at 1558 nm. An amplitude modulator (AM) synchronized with the swept pump modulated the CW TLS into 360-ps pulses, which was then amplified by two EDFAs as the quasi-CW pump. A 0.7-nm bandwidth filter was deployed between the two EDFAs to suppress the amplified spontaneous emission (ASE) noise. The two pumps were then combined through a wavelength-division multiplexer (WDM) and temporally overlapped using an optical tunable delay line. The SUT was a sub-ps pulse with 5-nm bandwidth at 1255 nm obtained through supercontinuum generation of the same MLFL, and it was thus optically synchronized with the two pumps. The SUT then propagated through 200-m of dispersion compensating fiber (DCF), which provided an input dispersion of 15 ps$^2$. Finally, the pumps and the SUT were combined using an O/C band WDM and then launched together into the 30-m HNLF with a nonlinear coefficient γ = 24 (W$^{-1}$km$^{-1}$) and a zero-dispersion wavelength of 1395 nm. The pumps and the HNLF formed a time lens, which induced quadratic phase modulation onto the SUT through FWM-BS. The peak power of the swept pump (P$_1$) and the quasi-CW pump (P$_2$) were adjusted such that P$_1$ = P$_2$, and (P$_1$+P$_2$)·γ·L = π in the HNLF to achieve the highest conversion efficiency during FWM-BS [20]. After the time lens, a narrow-band idler generated through FWM-BS was filtered out before it propagated through 2 DCF modules that provide a total output dispersion of 1958 ps$^2$. Overall, the system functions as a temporal magnifier, and the output is a temporally magnified SUT, which would then be characterized by the subsequent TCSPC system that consisted of a near-infrared SPAD and timing electronics (See Supplement 1 Section 1 for detailed set-up).

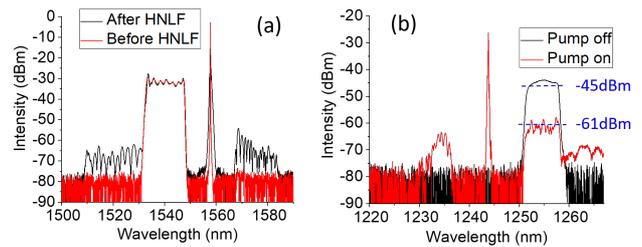

Fig. 3. Optical spectra of FWM-BS at the pump (a) and the signal (b) bands. The conversion efficiency of 97% was achieved, and it could be further improved towards unity by optimizing the spectral flatness of the swept pump.

The corresponding optical spectrum is shown in Fig. 3. Owing to the large spectral separation between the signal and the pump, the spectra were measured at the pump band (Fig. 3(a)) and signal band (Fig. 3(b)) individually. As shown in Fig. 3(a), the pumps consisted of a 14-nm swept pump centering at 1540 nm and a quasi-CW pump at 1558 nm. The spectral modulations of the swept pump are inherited from the spectrum of the MLFL. After passing through the HNLF, some inevitable parasitic FWM processes happened between the two pumps. However, their influence was negligible because the conversion efficiency is less than 0.1% and they are spectrally separated from the SUT. As shown in Fig. 3 (b), when the two pumps were turned on, the 5-nm signal at 1255 nm was converted to a narrow band idler at 1244 nm with a close-to-unity conversion efficiency, and the frequency separation between the signal and idler equals the frequency difference between the two pumps. Comparing with the initial signal, more than 16 dB signal depletion was achieved. Similar to the pump side, some parasitic FWM could be observed. However, they only consisted of 0.76% of the original signal power. Therefore, 97% of the signal was efficiently converted to the narrow band idler, and the efficiency could be further improved towards unity by optimizing the spectral flatness of the swept pump. A free-space grating monochromator was used to filter out the 1244-nm idler with a 100 dB extinction ratio.

In the TCSPC that characterizes the temporally magnified SUT, the SPAD was operated at gated mode with a gate-on time of 5 ns and a gate frequency of 25 MHz synchronized with the SUT. The hold-off time for the SPAD was set to 5 µs to suppress afterpulsing. Under such settings, the dark count was found to be 6,000/s. The input to the SPAD was then attenuated properly such that the maximum detection probability per gate was 1% to minimize the pile-up effect. Therefore, the maximum counting rate was about 110,000/s. Overall, the performance of the TM-TCSPC is summarized in Fig. 4. Fig. 4(a) shows the TM-TCSPC timing histograms obtained with different start-stop times. The FWM-BS conversion efficiency gradually rolled off as the start-stop time increases and the 10-dB record length was measured to be 30 ps. It consisted of only 18% of the available 165-ps aperture of the time lens owing to the restriction of phase matching. Therefore, the record length can be further expanded by using HNLF with a lower dispersion slope. A zoom-in plot of the central histogram in (a) is shown in the inset of Fig. 4(b) with a full-width at half maximum of 123 ps. Fig. 4(b) plots the magnified start-stop time with respect to the start-stop time within the 30-ps record length, showing a linear relationship with a slope of 130 that is the temporal magnification ratio. To demonstrate the potential of TM-TCSPC in ultrafast fluorescence lifetime measurement, we analyzed its capability to resolve small pulsewidth changes of about 130 fs (See supplement 1 section 2). SUTs with four different pulsewidths were first calibrated with a background-free second harmonic generation intensity autocorrelator (AC) and then measured using the TM-TCSPC (Fig. S2(b) and Fig. S2(c) respectively). Evidently, the TM-TCSPC can differentiate all four SUTs with different pulsewidths. An unprecedentedly short SPTR of 550 fs was estimated by fitting these measurements (See supplement 1 section 2 for details). Fig. 4(c) plots the TM-TCSPC-measured pulsewidths (red stars) overlaid with the AC-calibrated pulsewidths (black stars), showing an excellent agreement with an r.m.s. fitting error of only 22 fs.

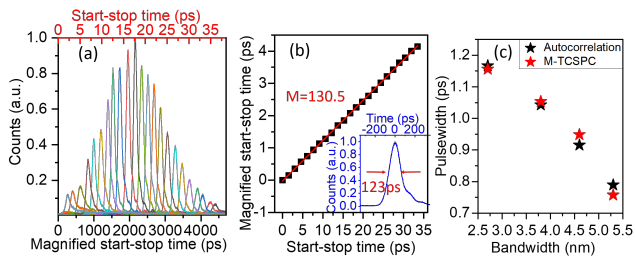

Fig. 4. Characterization of TM-TCSPC (a) Measured output photon counting histograms as the input pulse is delayed. (b) Temporal shifting of the histogram peak as a function of input time extracted from (a), featuring a magnification ratio of 130. Inset: zoom-in of the central histogram in (a) showing a full-width at half maximum of 123 ps. (c) Measurement deviations between autocorrelation and TM-TCSPC.

At zero start-stop time, the sensitivity of the system was also characterized and the minimum measurable SUT power was -67 dBm, corresponding to about 20 photons per pulse. The sensitivity is currently limited by the dark count of the SPAD and the large insertion loss of the output DCF modules (32 dB). Detection sensitivity of -95 dBm (0.03 photons per pulse) was measured before the DCF modules, which was then limited by the spontaneous Raman scattering noise. Therefore, by replacing the DCF modules with a low-loss chirped fiber Bragg grating (CFBG), the sensitivity of the system can be significantly enhanced by 28 dB, allowing efficient processing and characterization of quantum-level SUT.

Finally, a proof-of-principle photon counting ToF 3D imaging was demonstrated to further highlight the benefit of sub-ps time resolution of TM-TCSPC, where its unprecedented 550-fs SPTR is translated to 82-µm depth resolutions in air. As shown in Fig. 5(a), the imaging sample consisted of four small pieces of glass with different heights glued onto a glass slide, which was then sputtered with chromium coating to eliminate multi-surface reflection. The heights of the four glass pieces were independently characterized to be 520 µm (upper left), 400 µm (upper right), 990 µm (lower left), and 180 µm (lower right) using spectral-domain optical coherence tomography (SD-OCT) [24].

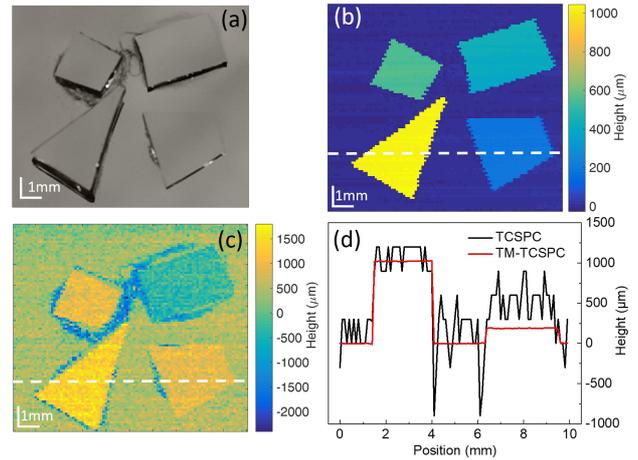

Fig. 5. Demonstration of ToF 3D imaging. (a) Photograph of the sample. (b) 3D image obtained using TM-TCSPC. (c) 3D image obtained using conventional TCSPC. (d) Comparison of the height profiles with TM-TCSPC (red trace) and conventional TCSPC (black trace) along the white dashed line. Evidently, TM-TCSPC offers orders-of-magnitude higher measurement precision and accuracy.

To acquire the ToF 3D image, the sample was put under a low-resolution confocal microscope with the TM-TCSPC attached to the return signal port. Each timing histogram was denoised by a lowpass filter before its peak is located and translated from time to depth (See Supplement 1 section 3). The resulting ToF 3D image is shown in Fig. 5(b). The heights of all four glass pieces were differentiated even with only 120-µm height difference between the upper two glass pieces. The measured heights were 565 µm (upper left), 382 µm (upper right), 1010 µm (lower left), and 182 µm (lower right), which matched well with the SD-OCT calibration results. The measurement accuracy, defined as the r.m.s. error between TM-TCSPC measurement and SD-OCT calibration, is calculated to be 26 µm. The limit of the measurement accuracy will be discussed in the next paragraph. Besides, the TM-TCSPC featured high measurement precision. As shown in the red trace of Fig. 5(d), the height measurement along the white dashed line shows a standard deviation of only 3 µm. As a comparison, the ToF 3D image obtained using conventional TCSPC is shown in Fig. 5(c) and the height variation along the same line as that in Fig. 5(b) is shown as the black trace in Fig. 5(d). Evidently, TM-TCSPC offers orders-of-magnitude higher measurement precision and accuracy over conventional TCSPC. More significantly, conventional TCSPC image shows much more evident cross-talk between depth and intensity information (see Supplement Fig. S3(a) for reference). Specifically shown in Fig. 5d, the measured height of the bottom right glass piece has a large error of 350 µm and the glass edges were even measured to exhibit negative heights that were unphysical. Such cross-talk stems from the photon pile-up effect in SPAD and is referred to as range walk error (RWE) in photon counting ToF 3D imaging systems [25]. As shown in Supplement 1 section 4, a 15-ps RWE was observed when the detection probability increased from 0.1 % to 1 %, a range typically used for TCSPC. Such a phenomenon is detrimental for high-resolution ToF 3D imaging because it not only results in mm scale depth error but also severely limits the intensity dynamic range. While RWE still exists in the TM-TSCPC, it is significantly suppressed by 99.2 % (130 times) thanks to the selective temporal magnification of the depth-induced timing. As a result, the RWE is reduced to 19 µm, which is in good agreement with the 26-µm measurement accuracy shown in Fig. 5.

In conclusion, we have demonstrated a TM-TCSPC that enables photon counting at the femtosecond regime for the first time. A temporal magnification of 130 with a 97% photon conversion efficiency has been achieved while maintaining the signal quantum coherence. Thus, the TM-TCSPC exhibits an unprecedentedly short SPTR of 550 fs. The TM-TCSPC can resolve ultrashort pulses with a 130-fs pulsewidth difference at a 22-fs accuracy. By replacing the DCF modules in the current system with a low-loss CFBG, detection sensitivity of -95 dBm (0.03 photons per pulse), limited only by the spontaneous Raman scattering noise, can be achieved. When applied to photon counting ToF 3D imaging, the TM-TCSPC greatly suppresses the RWE that limits all photon counting ToF 3D imaging systems by 99.2 % (130 times) and thus provides high depth measurement accuracy and precision of 26 µm and 3 µm, respectively. In combination with computer vision [26], such unprecedentedly high depth resolution will enable human face recognition even when the object is beyond line-of-sight [27, 28]. The TM-TCSPC is a promising solution for photon counting at the femtosecond regime and it will also benefit various other research fields such as low-light fluorescence lifetime spectroscopy and microscopy, time-gated Raman spectroscopy, light-in-flight imaging, and computational diffuse optical tomography.

**Funding.** National Science Foundation (NSF) (1021188); Office of Naval Research (ONR) (N00014-19-1-2251).

**Disclosures.** The authors declare no conflicts of interest.

See Supplement 1 for supporting content.

# Time-magnified photon counting with a 550-fs resolution: supplementary material

BOWEN LI,[1] JAN BARTOS, YIJUN XIE, AND SHU-WEI HUANG[*]

*Department of Electrical, Computer, and Energy Engineering, University of Colorado, Boulder, Colorado, US*
[1] *Email: Bowen.Li@colorado.edu*
*Corresponding author: Shu-Wei.Huang@colorado.edu*

This document provides supplementary information to "Time-magnified photon counting with a 550-fs resolution," It consists of four sections: 1. TM-TCSPC schematic diagram; 2. Resolving sub-ps pulsewidth change with TM-TCSPC; 3. ToF image and signal processing; and 4. Photon pile-up effect.

## 1. TM-TCSPC schematic diagram

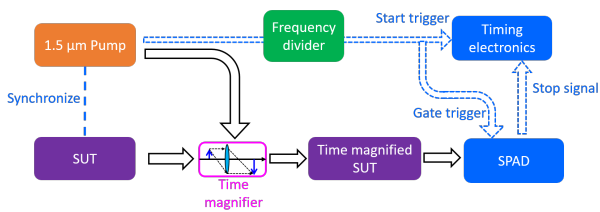

Fig. S1. The schematic diagram for the TM-TCSPC. SUT, signal under test; SPAD, single-photon avalanche diodes.

Since the set-up schematic in Fig. 2 focuses on the fiber parametric temporal magnifier, more details on TM-TCSPC measurement configuration are provided in Fig. S1. Since a 100 MHz pump source was used for both pumping the temporal magnifier and generating the 1.2-μm signal under test (SUT), the output signal from the temporal magnifier also had a repetition rate of 100 MHz. However, the SPAD used in the experiment (MPD, PDM-IR) only supported a maximum trigger frequency of 25 MHz when operating at gated mode. To synchronize the gate window of the SPAD with the optical input signal, 1 % of the 1.5-μm source was received by a photodetector to generate a 100 MHz electrical pulse train, which was then frequency divided by 4 times using an electrical frequency divider to generate a 25-MHz electrical pulse train as the trigger. This trigger was then split equally into two parts. One part served as the gate trigger to synchronize the gate window in the SPAD, while the other part was used as the start trigger for the timing electronics module (Hydraharp 400). Once the SPAD detects a photon, a stop signal would be generated and sent to the timing electronics module, registering the photon in the time bin. The instrument response function (IRF) of the whole TCSPC system was measured to be 90 ps using fs pulses.

## 2. Resolving sub-ps pulsewidth change with TM-TCSPC

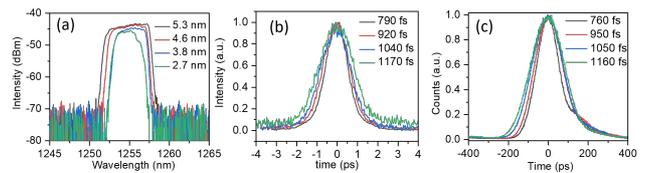

Fig. S2. Resolving sub-ps pulsewidth change with TM-TCSPC. (a) Optical spectra of the input pulses with different bandwidths. (b) Background-free SHG intensity autocorrelation traces of the corresponding pulses. Legends show the measured pulsewidth after deconvolution (c) TM-TCSPC timing histograms of the corresponding pulses. Legends show the measured pulsewidth after demagnification and deconvolution.

As shown in Fig. S2(a), the different pulsewidths used in Fig. 4 were obtained by varying the optical bandwidth using a variable bandpass filter. Four different bandwidths: 2.7 nm, 3.8 nm, 4.6 nm, and 5.3 nm were used for the test. Before the direct measurements using TM-TCSPC, pulsewidths were first characterized using a background-free second harmonic generation (SHG) intensity autocorrelator as references and the autocorrelation traces are shown in Fig. S2(b). The measured pulsewidths after deconvolution were 790 fs, 920 fs, 1040 fs, and 1170 fs respectively, and the change of pulsewidth in each step was only around 130 fs. Then, the pulses were measured again using the TM-TCSPC with the timing histograms shown in Fig. S2 (c). To obtain the pulsewidth, the TM-TCSPC results needed to be demagnified by the temporal magnification ratio and then deconvolved with the system IRF. The fitted IRF that minimizes the r.m.s. difference between the pulsewidths measured by the two methods was calculated to be 550 fs. Based on the fitted IRF, the measured pulsewidth using TM-TCSPC were 760 fs, 950 fs, 1050 fs, and 1160 fs, respectively. Of note, the dispersion slope in the output dispersion will induce imaging aberration of the temporal magnifier, resulting in IRF distortion that depends on the SUT bandwidth. Such influence can be observed from the black trace in Fig. S2(c), where the pulse pedestal is deviated from a Gaussian shape owing to the dispersion slope.

### 3. ToF image and signal processing

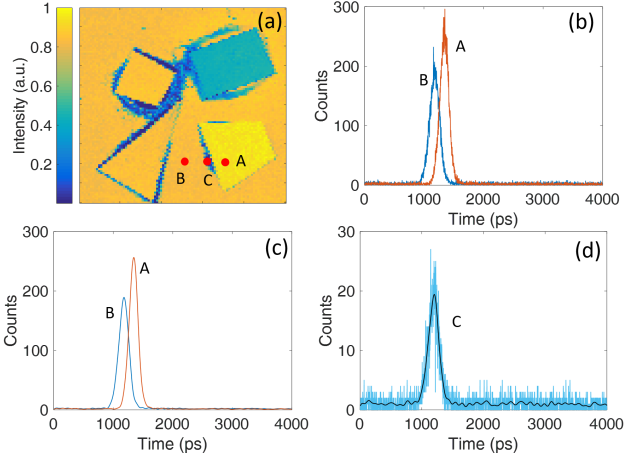

Fig. S3. (a) Intensity image of the glass sample. (b) Original timing histogram obtained by TM-TCSPC system at two different positions marked on (a). (c) Processed timing histogram using low-pass filtering in the frequency domain. (d) Original and processed timing histograms at point C in (a).

The sample was mounted on a motorized 2D translation stage and was scanned across a 10 mm by 10 mm area through 10,000 steps for imaging. 1.2-μm signal with 5-nm bandwidth was focused onto the sample. The intensity image shown in Fig. S3(a) was obtained from the peak height of the photon-histogram using conventional TSCPC. The different intensities observed at each glass sample were due to the limited working distance of the focusing lens. Lower reflected power was collected for surfaces that are slightly out of focus. Besides, near-zero intensities were measured at few pixels on the glass edges, owing to the tilted surface. Other low-intensity areas outside the glass edges were caused by the coating defects. However, the ToF depth imaging should not be influenced by the intensity variation since only height contrasts are measured. The raw data obtained using TM-TCSPC at two different points A and B marked by red dots in Fig. S3(a) is shown by the orange and blue trace in Fig. S3(b), respectively. Both timing histograms were acquired with an integration time of 0.3 s and the bin size for the histograms was 2 ps. As can be observed, the two timing histograms are well separated. To further increase the accuracy in determining the peak location, both traces were filtered in the frequency domain with a bandwidth of 10 GHz. The filtered traces are shown in Fig. S3(c), which show a much higher signal-to-noise ratio (SNR). According to Fig. S3(c), temporal separation of 166 ps was found between the two histograms, which correspond to a height difference of 190 μm. The histogram for point C where the defective coating induces a 10 dB less reflectivity is shown in Fig. S3(d). Even though the original histogram exhibits much worse SNR owing to a lower number of photon counts, the black trace after filtering still shows sufficient SNR to precisely locate the peak location.

### 4. Photon pile-up effect

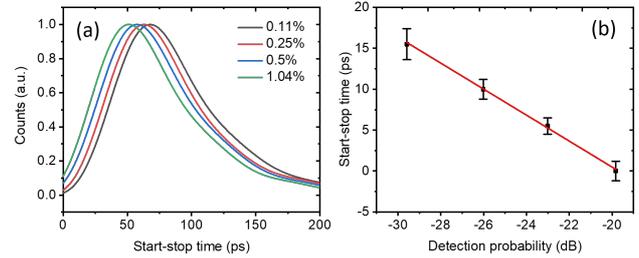

Fig. S4. Investigation of pile-up effect on the error of the TCSPC timing histogram. (a) Four timing histograms with different detection probability. (b) Peak locations of the timing histograms with respect to the detection probability. Linear fitting shows a peak shift of -1.6 ps/dB. Four repeated measurements were performed for each detection probability to obtain the mean value and standard deviation.

The large depth measurement error observed in Fig. 5(c) is attributed to the photon pile-up effect. To further confirm our hypothesis, an experiment was conducted by launching quantum level femtosecond pulses at 1.5 μm directly into the SPAD. A continuously variable metallic neutral density filter (Thorlabs, NDC-25C-2) was used to change the detection probability while all other experimental conditions were kept the same. A longer integration time of 10 s was used to further increase the SNR to guarantee the measurement accuracy. Four different detection probabilities were used for the test: 0.11%, 0.25%, 0.50% and 1.04%. The corresponding TCSPC timing histograms are shown in Fig. S4(a). Obvious timing shifts are observed at different detection probabilities. For each detection probability, four repeated measurements were performed to obtain the mean and standard deviation of the peak locations of the TCSPC timing histograms. The results are shown in Fig. S4(b). A monotonic timing shift towards shorter time can be observed with increasing detection probability. Around 15-ps timing shift was introduced when the detection probability was increased from 0.11 % to 1.04 %. The slope from the linear fitting is -1.6 ps/dB. Therefore, we confirm that photon pile-up effect can induce picosecond scale shift on the TCSPC timing histograms even when the detection probability is below 1%, which is detrimental for mm and sub-mm scale ToF depth imaging. From this perspective, the mitigation effect from the TM-TCSPC technique is important for achieving accurate measurement results.